\documentclass[Proceedings]{SciPost2}
\usepackage{graphicx}
\usepackage{subcaption}

\usepackage{url}
\usepackage{hyperref}
\definecolor{nicered}{rgb}{0.5,0.,0.}
\definecolor{nicegreen}{rgb}{0.,0.5,0.}
\definecolor{niceblue}{rgb}{0.,0.,0.5}
\hypersetup{colorlinks,citecolor=nicegreen,linkcolor=nicered,urlcolor=niceblue}

\usepackage{amsmath,amssymb}
\usepackage{physics}
\DeclareSymbolFont{usualmathcal}{OMS}{cmsy}{m}{n}
\DeclareSymbolFontAlphabet{\mathcal}{usualmathcal}

\newcommand{\bea}[1]{\begin{equation}
	\begin{aligned}#1
	\end{aligned}
	\end{equation}}

\begin{document}

{\hfill FERMILAB-CONF-21-327-T, PITT-PACC-2115, SMU-HEP-21-08}

\begin{center}{\Large \textbf{
A general-mass scheme for prompt charm production at hadron colliders}
}\end{center}

\begin{center}
Keping Xie\textsuperscript{1$\star$},
John M. Campbell\textsuperscript{2},
and Pavel M. Nadolsky\textsuperscript{3}
\end{center}

\begin{center}
{\bf 1} Pittsburgh Particle Physics Astrophysics and Cosmology Center,
Department of Physics and Astronomy, University of Pittsburgh, Pittsburgh, PA 15260, USA\\
{\bf 2} Theoretical Physics Department, Fermilab, Batavia, IL 60510, USA\\
{\bf 3} Department of Physics, Southern Methodist University, Dallas, TX 75275-0175, USA\\
$\star$xiekeping@pitt.edu
\end{center}

\begin{center}
\today
\end{center}

\pagestyle{fancy}
\fancyhead[LO]{\colorbox{scipostdeepblue}{\strut \bf \color{white}~Proceedings}}
\fancyhead[RO]{\colorbox{scipostdeepblue}{\strut \bf \color{white}~DIS 2021}}


\definecolor{palegray}{gray}{0.95}
\begin{center}
\colorbox{palegray}{
  \begin{tabular}{rr}
  \begin{minipage}{0.1\textwidth}
    \includegraphics[width=22mm]{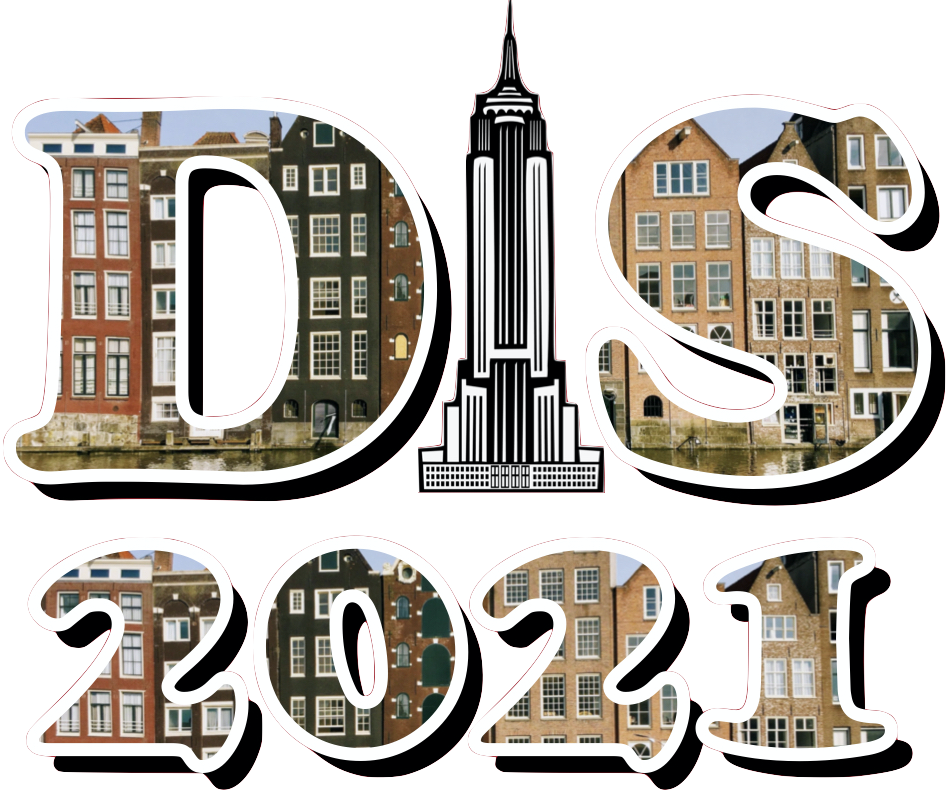}
  \end{minipage}
  &
  \begin{minipage}{0.75\textwidth}
    \begin{center}
    {\it Proceedings for the XXVIII International Workshop\\ on Deep-Inelastic Scattering and
Related Subjects,}\\
    {\it Stony Brook University, New York, USA, 12-16 April 2021} \\
    \doi{10.21468/SciPostPhysProc.xx.xxxx}
    \end{center}
  \end{minipage}
\end{tabular}
}
\end{center}

\section*{Abstract}
{
In these proceedings, we apply the recently developed S-ACOT-MPS factorization scheme at the next-to-leading order to prompt charm production at hadron colliders. It provides a good agreement with experimental data on charm meson production measured by LHCb at 7 and 13 TeV. The low-$p_T$ data are on the margins of the theoretical error bands, emphasizing the importance of including contributions beyond the next-to-leading order.
}


\section{General-mass calculations for heavy-flavor hadroproduction}
\label{sec:intro}
Heavy-flavor hadrons provide an excellent venue to test our understanding of the strong interaction. The most effective way to produce large numbers of heavy-flavor hadrons is through hadronic collisions, due to the large cross sections. Thanks to the sufficiently large heavy-quark mass, heavy-flavor hadroproduction is perturbatively calculable in Quantum Chromodynamics (QCD). In the last few decades, numerous advancements have been made in the understanding of charm and bottom production at the Tevatron and Large Hadron Collider (LHC). See Ref.~\cite{Frixione:1997ma} for a review of heavy-quark phenomenology at hadron colliders.

However, predicting  heavy-flavor production can be complicated, as dominant production mechanisms vary depending on the energy regime being probed. When a heavy quark's transverse momentum is of order of the quark mass or less, \emph{i.e.}, $p_T^2\lesssim m_Q^2$,\footnote{This criterion is equivalent to having partonic energy around the threshold, $\widehat{s}\gtrsim 4m_Q^2$.} heavy quarks are mainly produced through flavor creation (FC), and hence, the fixed-flavor-number (FFN) scheme \cite{Nason:1987xz,Nason:1989zy,Beenakker:1988bq,Beenakker:1990maa} can give a good description. In this approach, the heavy quark is only included in the hard scattering processes, with the $(p_T^2/m_Q^2)^n$ terms  kept in the perturbative coefficients. In contrast, when the transverse momentum becomes much larger than the quark mass ($p_T^2\gg m_Q^2$), the FFN scheme eventually becomes unreliable, as the missing high-order terms grow as $\alpha_s^m\log^n(p_T^2/m_Q^2)$. As a solution, the large logarithms are resummed in heavy-flavor parton distribution functions or fragmentation functions. In such a scenario, the heavy quark mass becomes negligible in the hard-scattering Feynman graphs. The zero-mass (ZM) approximation of the variable-flavor number (VFN) scheme then applies, treating the heavy quark as an active parton in the evolution of $\alpha_s$ and PDFs, and with the heavy flavor mainly produced through flavor excitation (FE). In the intermediate energy region, $p_T^2\sim m_Q^2$, we need a composite scheme to bridge the FFN to ZM schemes, called a general-mass variable flavor number (GM-VFN) scheme.

Several GM-VFN schemes, such as the ACOT \cite{Aivazis:1993kh,Aivazis:1993pi}, KKSS~\cite{Kniehl:2004fy,Kniehl:2011bk}, RT~\cite{Thorne:1997ga} and FONLL \cite{Cacciari:1998it,Cacciari:2001td}, are now being used for determination of PDFs \cite{Hou:2019efy,NNPDF:2017mvq,Bailey:2020ooq}; some have been successfully applied to the processes of heavy-flavor production  at the Tevatron~\cite{Kniehl:2004fy} and LHC~\cite{Kniehl:2012ti,Cacciari:2012ny}.
Recently, by following the idea of the rescaling variable $\chi$ in the DIS simplified-ACOT-$\chi$ scheme \cite{Tung:2001mv}, Helenuis and Paukkenen adopted a kinematic ($m_{T}$) dependent momentum fraction to describe $D$-meson production at LHCb and ALICE \cite{Helenius:2018uul}.  In our own work along this line, we introduce a massive phase space (MPS) for the final-state heavy flavors, which allows us to calculate the production rate in a fully differential way~\cite{Xie:2019eoe}. We summarize some key results here and present the full formalism in \cite{Xie:2019eoe} and an upcoming publication. 

\section{The simplified SACOT scheme with massive phase space}
\label{sec:theory}
In this section, we will introduce the general framework of our method -- ``the simplified ACOT scheme with massive phase space", shortened as S-ACOT-MPS \cite{Xie:2019eoe}. 

As we mentioned in Sec.~\ref{sec:intro}, heavy flavors can be produced either through flavor creation or flavor excitation mechanisms. However, the diagrams with collinear splitting of gluon or light quarks into heavy-quark pairs are counted both in the FC and FE terms. To avoid the double counting we have to subtract the overlap contribution, which is called the subtraction (SB) or asymptoti term \cite{Tung:2001mv}. In the ACOT scheme, the subtraction is constructed from the FE partonic cross section convoluted with the light-to-heavy collinear splitting arising in the counterpart FC coefficients. The total cross section can be constructed as
\begin{equation}\label{eq:ACOT}
\sigma_{\rm ACOT}=\sigma_{\rm FC}+\sigma_{\rm FE}-\sigma_{\rm SB}.
\end{equation}
In the high energy limit, $\widehat{s}\gg 4m_Q^2$, $\sigma_{\rm SB}$ is expected to cancel the corresponding logarithmic contributions from $\sigma_{\rm FC}$,\footnote{Still, a finite part of $\sigma_{\rm FC}-\sigma_{\rm SB}$ will remain.} and the ZM approximation adopted in the FE term becomes valid. 
Conversely, around the threshold region, $\widehat{s}\gtrsim 4m_Q^2$ or $p_T^2\lesssim m_Q^2$, the contributions from SB and FE terms cancel, reducing  $\sigma_{\rm ACOT}$ to the FC term. 

In inclusive heavy-flavor hadroproduction, $pp\to QX$, the components in Eq.~(\ref{eq:ACOT}) up to the next-to-leading order (NLO) can be written as 
\bea{
&\sigma_{\rm FC}=\sum_{i,j}f_{i}(x_i,\mu^2)f_{j}(x_j,\mu^2)\widehat{\sigma}_{ij\to QX},\\
&\sigma_{\rm FE}=\sum_{i}f_{i}(x_i,\mu^2)f_{Q}(x_Q,\mu^2)\widehat{\sigma}_{iQ\to QX}+(i\leftrightarrow Q),\\
&\sigma_{\rm SB}=\sum_{i,j}f_{i}(x_i,\mu^2)\left[P_{Qj}\otimes f_{j}\right](x_Q,\mu^2)\widehat{\sigma}_{iQ\to QX}
+(i\leftrightarrow Q),\\
}
where $i,j$ denote the light flavor, including gluon and quarks. The hard matrix elements of the FE terms are the same as the SB ones. We compute them by setting $m_Q =0$, and we convolve the SB term with a ``subtraction" PDF for the heavy flavor, 
\begin{equation}
\tilde{f}_{Q}(x,\mu^2)=\sum_j\left[P_{Qj}\otimes f_{j}\right](x_Q,\mu^2),
\end{equation}
where the logarithmic terms are implicitly included in during the convolution. 
At the leading order,\footnote{This corresponding to the NLO to $pp\to QX$, as the leading order purely comes from FC, $gg\to Q\bar{Q}$.} only heavy quark from gluon splitting, $\alpha_s LP_{Qg}$, contributes, where $L=\log(\mu^2/m_Q^2)$. At one order higher, both double splittings, $(\alpha_s LP)^2$, and higher-order corrections, $\alpha_s^2LP^{(2)}$, will participate. In such a way, the subtraction framework of the ACOT scheme in Eq. (\ref{eq:ACOT}) is expected to work order by order.

At $\widehat s \to 4m_Q^2$, the partonic cross section $\widehat{\sigma}_{iQ\to QX}$ is soft and collinearly divergent if all mass terms are neglected,  which makes the difference of the FE and SB terms perturbatively unstable. As a generalization of the $\chi$ prescription in DIS at the mass threshold \cite{Tung:2001mv}, we integrate all terms -- FC, FE, and SB -- using massive phase space (MPS), which enforces energy-momentum conservation at the threshold and regulates both soft and collinear divergences. As a result, the FE and SB cancellation is stabilized, which ensures the simplified-ACOT scheme to smoothly match to the FFN one in the low-energy limit. We name this novel approach as ``S-ACOT-MPS'' scheme~\cite{Xie:2019eoe}. 

\begin{figure}\centering
\includegraphics[width=0.48\textwidth]{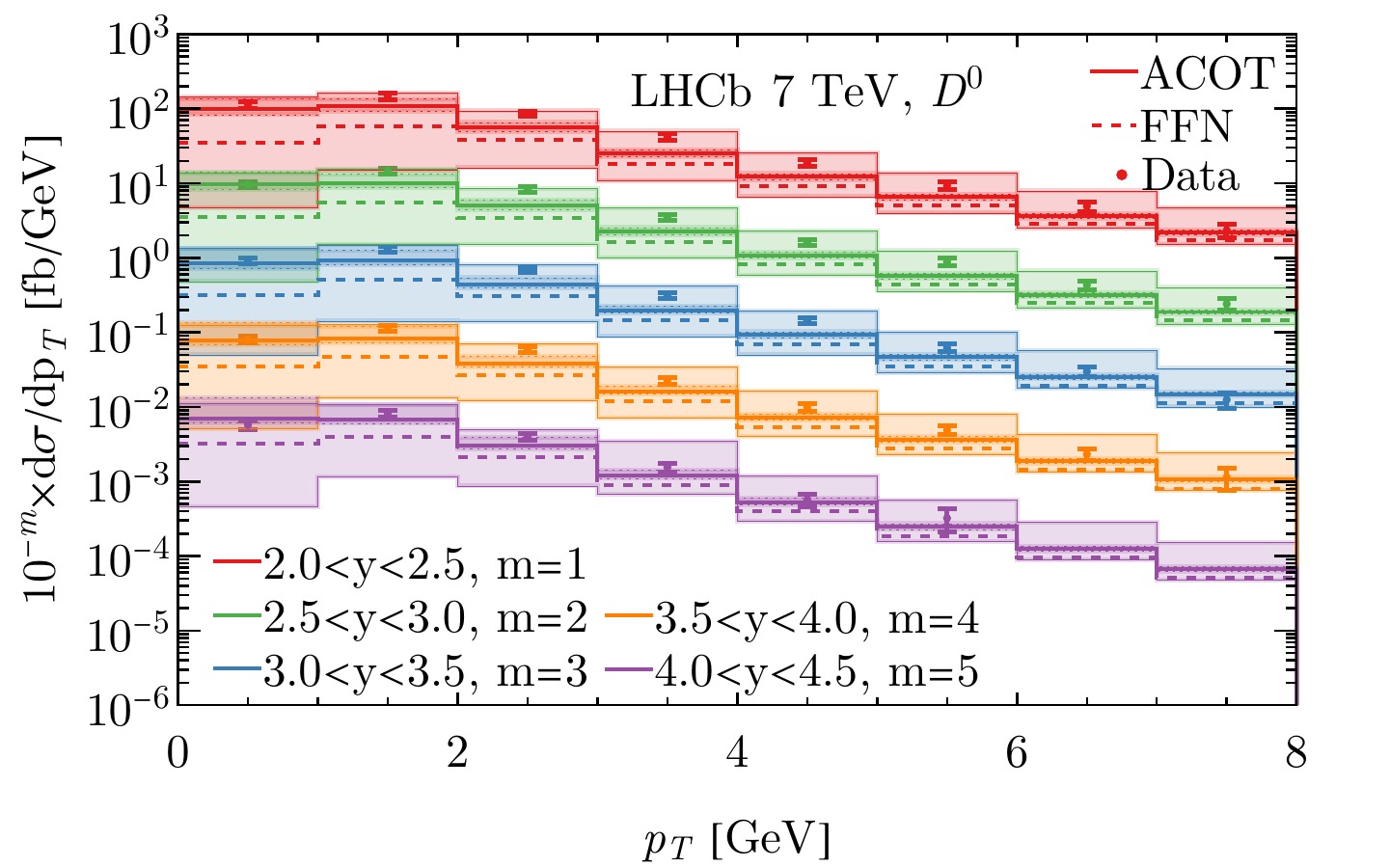}
\includegraphics[width=0.48\textwidth]{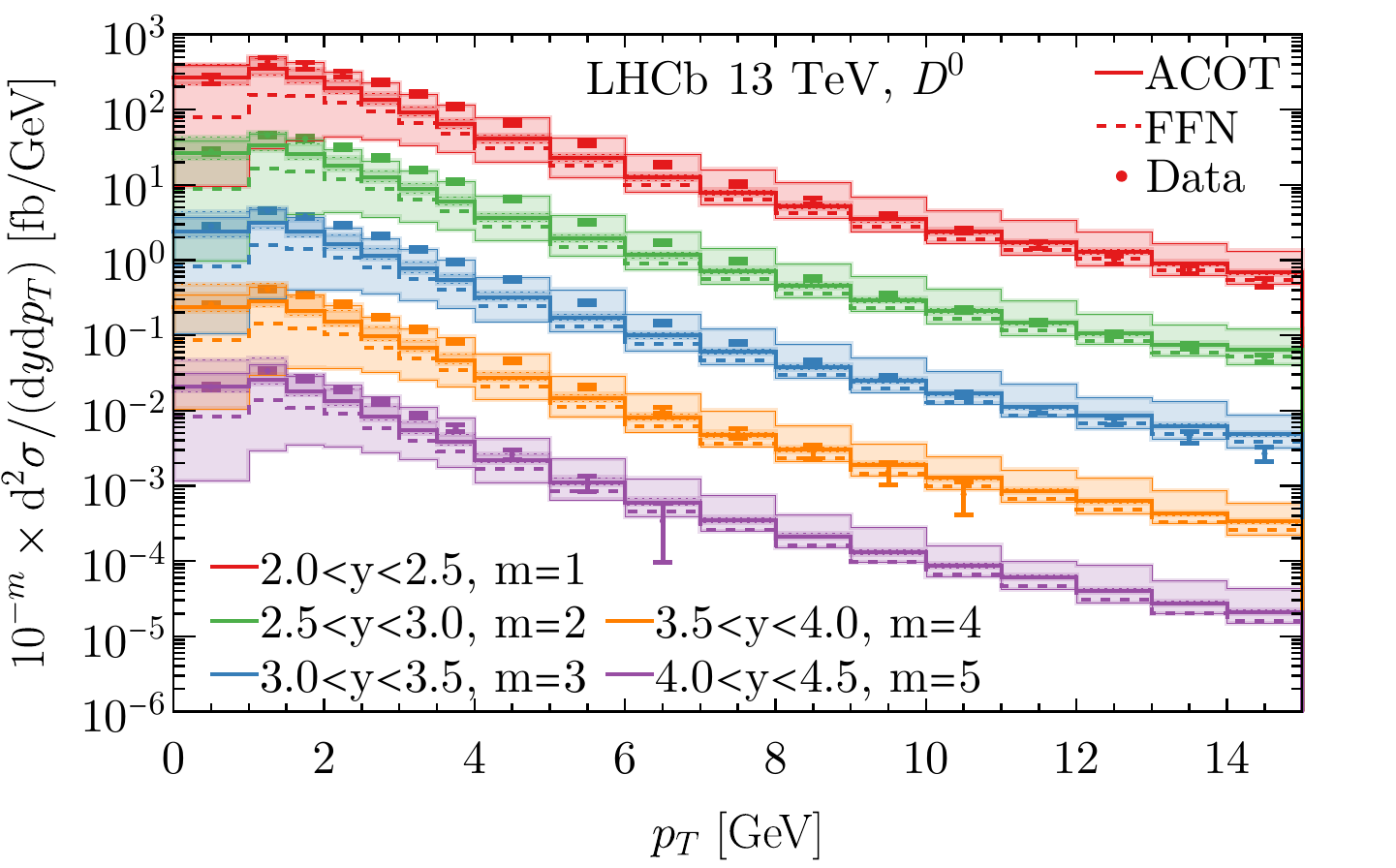}	
\caption{The differential cross sections of $D^0$ meson production in the rapidity region $2.0<y<4.5$ compared with LHCb 7 and 13 TeV data \cite{Aaij:2013mga,Aaij:2015bpa}. The larger error bands indicate the dominating scale uncertainties, while smaller dotted ones are the sub-leading PDF uncertainties.}
\label{fig:LHCbD0}
\end{figure}

Equipped with this new technique, we predict the differential cross sections of $D^{0}$ production, shown in Fig. \ref{fig:LHCbD0}. 
We choose the CT18NLO PDF~\cite{Hou:2019efy} in this example, which matches to the order of the perturbative calculation.
The fragmentation of a charm quark into a $D^0$ meson is parameterized as a transition probability $f(c\to D^0)=0.565\pm0.032$, measured at $e^+e^-$ colliders \cite{ParticleDataGroup:2008zun}. 
In order to explore the scale variation, we choose a default scale as 
\begin{equation}
\mu_R=\mu_F=\mu_0=\sqrt{p_T^2+(2m_c)^2},
\end{equation}
and vary $\mu_R$ and $\mu_F$ independently by a factor of two. Unlike a conventional choice, $M_T=\sqrt{p_T^2+m_c^2}$, this modified scale ensures that even the lowest choice of scale ($\mu_0/2$) satisfies $\mu_F \geq m_c$.  This means that the scale remains above the charm matching point, thus ensuring positivity of the charm PDF. 
When compared with the LHCb measurements at 7 TeV~\cite{Aaij:2013mga} and 13 TeV~\cite{Aaij:2015bpa}, our theoretical calculations provide good agreement with data within uncertainties. The low $p_T$ data are on the margins of the theoretical error bands, indicating the importance of missing higher-order (NNLO) contributions.

\section{Conclusion}
In this proceedings, we apply the recently developed Simplified ACOT scheme with Massive Phase Space (S-ACOT-MPS)~\cite{Xie:2019eoe} to charmed meson production at hadron colliders. The open charm diagrams are categorized  into Flavor Excitation (FE) and Flavor Creation (FC) terms, with overlapping graphs subtracted after their convolution with the collinear splitting of a gluon into a heavy-quark pair. The massive phase space is used with the FC, FE, and subtracted (SB) terms and accounts for the threshold effects of massive heavy quarks, which stabilizes the perturbative convergence around the threshold region. With a simple fragmentation fraction model, our prediction provides a good description of $D^0$ meson production measured in the LHCb detector at 7 TeV~\cite{Aaij:2013mga} and 13 TeV~\cite{Aaij:2015bpa}.

We publicly provide the subtracted and residual charm PDFs, which allow one to straightforwardly extend the S-ACOT-MPS general-mass scheme to other NLO calculations with heavy quarks at hadron colliders. Namely, with these PDFs one can compute at once the flavor-excitation and subtraction terms for charm-flavor initiated processes. The specific demonstration for the application in the framework of \texttt{MCFM}~\cite{Campbell:2019dru}, and fast \texttt{APPLgrid} computation tables can be obtained at \textrm{HEPForge}:
\url{https://sacotmps.hepforge.org}.

\section*{Acknowledgements}
We thank Marco Guzzi for insightful discussions. 
The work of KX at University of Pittsburgh is supported in part by
 the Department of Energy under Grant No. DE-FG02-95ER40896, the National Science Foundation under Grant No. PHY-1820760, and in part by PITT PACC. The work of PMN at SMU is supported by the Department of Energy under Grant No. DE-SC0010129.
The work of JMC at Fermilab was supported by Fermi Research Alliance, LLC under Contract No. DE-AC02-07CH11359 with the U.S. Department of Energy, Office of Science, Office of High Energy Physics.

\bibliography{ref.bib}

\end{document}